# Surface activity of cancer cells: the fusion of two cell aggregates

Ivana Pajic-Lijakovic*, Milan Milivojevic

Faculty of Technology and Metallurgy, Belgrade University, Serbia

iva@tmf.bg.ac.rs

**Abstract**

A key feature that distinguishes cancer cells from all other cells is the capability to spread throughout the body. Although a good comprehension of how cancer cells collectively migrate by following molecular rules which influence the state of cell-cell adhesion contacts has been generated, the impact of collective migration on cellular rearrangement from subcellular to supracellular level remains less understood. Thus, considering collective cell migration (CCM) of cancer mesenchymal cells on one side and healthy epithelial cells on the other during the fusion of two cell aggregates could result in a powerful tool in order to address the contribution of structural changes at subcellular level which influence the cellular rearrangements and help to understand this important, but still controversial topic. While healthy epithelial cells undergo volumetric cell rearrangement driven by the tissue surface tension, which results in a collision of opposite directed velocity front near the contact point between two cell aggregates, mesenchymal cells follow quite different scenario. These cells are capable of reducing the surface tension and undergo surface cell rearrangement. The main goal of this contribution is to discuss the origin of surface activity of cancer cells by accounting for the crosstalk between cell-cell and cell-ECM adhesion contacts influenced by the cell contractility.

**Introduction**

Cancer invasion through the extracellular matrix and the surrounding tissue is a key step in cancer disease progression (Clark and Vignjevic, 2015; Gandalovičová et al., 2016; Beunk et al., 2019; Kubitschke et al., 2021). Therefore, finding the correlation between CCM and cell rearrangement is the one of main and most challenging tasks in biomedical sciences. The configuration of migrating cells and the rate of its change during invasion induce the cell residual stress accumulation which has a feedback impact on: (1) the strength of cell-cell and cell-extracellular matrix (ECM) adhesion contacts (Liu et al, 2006; Devanny et al., 2021; Tian et al., 2021), (2) intracellular signaling cascades (Petrungaro et al., 2019), (3) the viscoelasticity of ECM (Pajic-Lijakovic and Milivojevic, 2020c; Tian et al., 2021), and (3) cell contractility in response to the microenvironmental conditions (Devanny et al., 2021). The ability of cancer cells to move in a specific tumor may be an indicator of a tumor's metastatic potential. However, despite extensive research devoted to study the movement of cancer cells, we still do not fully understand the phenomenon. CCM influences the viscoelasticity of multicellular systems which has the feedback impact on cell long-time rearrangement (Barriga and Mayor, 2019; Pajic-Lijakovic and Milivojevic 2019a,b; Blanchard et al., 2019).

The CCM is guided by the surface tension force, viscoelastic force and traction force (Pajic-Lijakovic and Milivojevic, 2020c). The surface tension force acts to reduce the surface of a multicellular system and induces cell movement from the aggregate surface region to the aggregate core region. This movement leads to an accumulation of cell residual stress within the cell aggregate core region. The phenomenon has been described by the Young-Laplace equation (Marmottant et al., 2009; Pajic-Lijakovic and Milivojevic, 2019c). An inhomogeneous distribution of cell normal residual stress represents a prerequisite for the generation of the viscoelastic force. It is resistive force acts always opposite to the direction of cell movement in order to suppress it (Pajic-Lijakovic and Milivojevic, 2020c). The traction force represents a consequence of interactions between cells and ECM and depends on the viscoelasticity of ECM (Murray et al., 1988).

The long-time rearrangement of mescenhymal cancer cell aggregates has been estimated and compared with the rearrangement of healthy epithelial cell aggregates by considering a simple model system such as the fusion of two cell aggregates (Kosztin et al., 2012; Shafiee et al., 2015; Dechristé et al., 2018; Grosser et al., 2021). The fusion of healthy epithelial cell aggregates, driven by the tissue surface tension, leads to a decrease in the surface of two-aggregate system, while the volume decreases or stays approximately constant (Shafiee et al., 2015; Grosser et al., 2021). The rearrangement of these systems follows two scenarios: (1) total coalescence (Shafiee et al., 2015) and (2) arrested coalescence (Oriola et al., 2020; Grosser et al., 2021). The arrested coalescence represents the consequence of the cell jamming state transition near the contact point between two cell aggregates (Oriola et al., 2020). The jamming state transition is caused by collision of velocity fronts near the contact point between two aggregates (Oriola et al., 2020; Grosser et al., 2021; Pajic-Lijakovic and Milivojevic, 2021b).

Surprisingly, the fusion of cancer cell aggregates follows quite different scenario by increasing the surface and volume of two-aggregate system. Dechristé et al. (2018) considered the fusion of two human carcinoma cell aggregates (HCT116 cell line) as a consequence of cell divisions within 70 h. The doubling time of HCT116 cells is 18 h (Gongora et al., 2008). Within this time period, cell divisions can be neglected, while the surface and volume changes occur primarily via CCM. The surface of two-aggregate system increase: (1) 3.47 times for larger cell aggregates (500 $\mu m$ diameter) and (2) 5.93 times for smaller aggregates (300 $\mu m$ diameter) within 18 h (Dechristé et al., 2018). Corresponding increase in the volume of two-aggregate system is: (1) 1.25 times for larger aggregates and (2) 1.55 times for smaller aggregates within 18 h (Dechristé et al., 2018). Grosser et al. (2021) considered and compared the fusion of two breast cell aggregates such as: (1) healthy epithelial MCF-10A cell lines and (2) cancerous mesenchymal MDA-MB-436 cells within 60 h. While MCF-10A cells undergo arrested coalescence and the average cell velocity drops to zero, cancer cells kept their velocity approximately constant within this time period and avoided the jamming state transition.

These findings point to a different scenario of cell rearrangement for cancer cells in comparison with the health epithelial cells. While health cells undergo volumetric rearrangement driven by the tissue surface tension (Shafiee et al., 2015; Grosser et al., 2021), cancer cells perform rather a surface rearrangement. Interestingly, cancer cells are able to reduce the tissue surface tension and behave as surface active constituents (Devanny et al., 2021). The main goal of this contribution is to emphasize the possible reason of this interesting phenomenon by accounting for the inter-relation between the state of cell-cell and cell-ECM adhesion contacts and cell contractility.

**The fusion of two cell aggregates: various scenarios**

The fusion of healthy cell aggregates is driven by the tissue surface tension which induces a decrease in the surface of cell-aggregate system while the volume decreases or stays approximately constant. The surface of two-aggregate system in contact was expressed as (Kosztin et al., 2012) $A(t) = 4\pi R(t)^2 (1 + cos\theta(t))$ while the total volume is (Dechristé et al., 2018): $V(t) = \frac{2\pi}{3} R(t)^3 (2 + 3cos\theta(t) - cos^3\theta(t))$ (where $R(t)$ is the aggregate radius and $\theta(t)$ is the fusion angle). The geometry of two-aggregate system is presented in Figure 1.

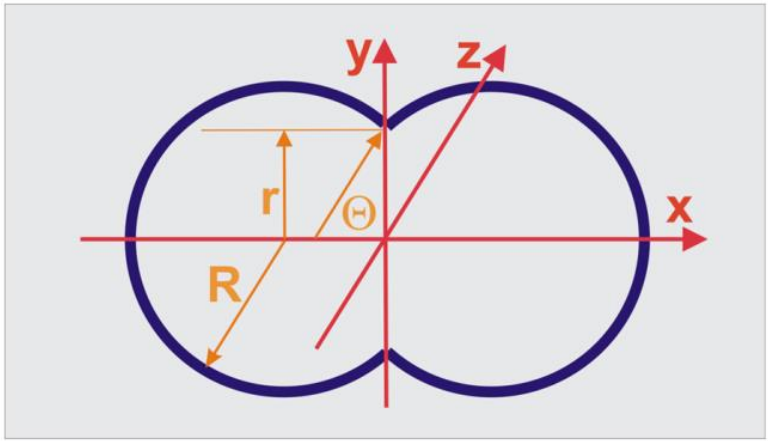

**Figure 1**. The geometry of two-aggregate system.

The aspect ratio is equal to $AR(t) = \frac{d(t)}{2r(t)}$ (where $d(t)$ is the longer axis equal to $d(t) = 2R(t)(1 + cos\theta)$ and $2r(t)$ is the shorter axis of multicellular system, while the neck radius is equal to $r(t) = R(t)sin\theta(t)$) (Dechristé et al., 2018).

Shafiee et al. (2015) considered the fusion of two confluent skin fibroblast cell aggregates and pointed out that the surface of the two-aggregate system decreases 2.18 times, while the volume decreases 2.38 times within 140 h (Figure 2a,b). The corresponding aspect ratio is $AR \approx 1.1$ after 140 h which points out that two-aggregate system reaches nearly spherical shape.

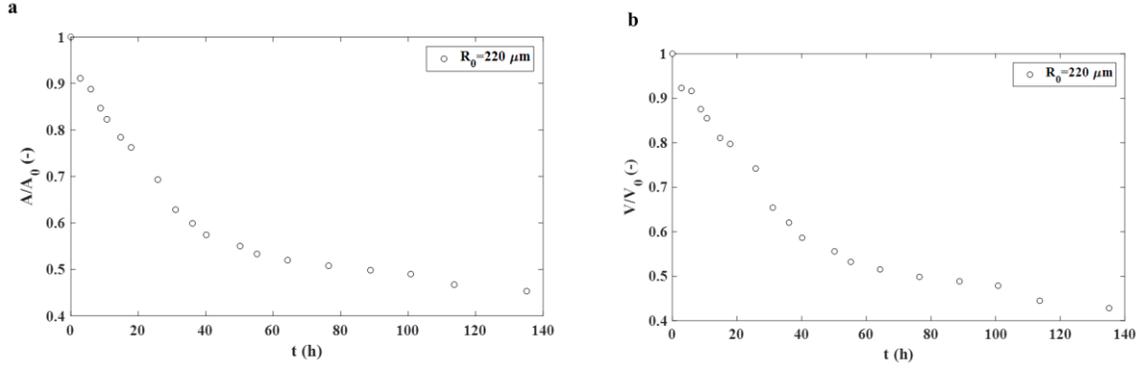

**Figure 2**. The fusion of two confluent skin fibroblast cell aggregates: (a) the surface of a two-aggregate system vs. time and (b) the volume of a two-aggregate system vs. time.

The surface and volume relax and reach the equilibrium states during the fusion process. Similar result is obtained for the fusion of two MCF-10A cell aggregates (Grosser et al., 2021). The decrease in the surface of two-aggregate system during the fusion is driven by the tissue surface tension. The surface tension guides the cell movement from the aggregate surface region to its core region by increasing in the cell normal residual stress within the core region. The phenomenon has been expressed by the Young-Laplace equation (Marmottant et al., 2009; Pajic-Lijakovic and Milivojevic, 2019c):

$$\gamma dA(t) = \sigma_i dV(t) \tag{1}$$

where $\gamma$ is the tissue surface tension and $\sigma_i$ is the internal cell normal residual stress generated during cell long-time rearrangement. The internal normal residual stress is equal to (Dechristé et al., 2018):

$$\sigma_i(t) = \left(\frac{\partial W_V}{\partial \varepsilon_V}\right) \tag{2}$$

where $\sigma_i = \sigma_i(\varepsilon_V)$, $\varepsilon_V = \frac{\Delta V}{V_0}$ is the volumetric strain equal to $\varepsilon_V(t) = (1 + \varepsilon_{xx})(1 + \varepsilon_{yy})(1 + \varepsilon_{zz}) - 1$, $\varepsilon_{xx}$, $\varepsilon_{yy}$, $\varepsilon_{zz}$ are the components of the volumetric strain, and $W_V(t)$ is the volumetric part of the strain energy density. If the volume and volumetric strain relax during the fusion process, the aggregate system can be treated as the viscoelastic solid (Pajic-Lijakovic, 2021). The corresponding long-time strain and volume relaxations caused by CCM follow exponential trends and on that base can be described by a linear viscoelastic model. The Zener model is the simplest linear constitutive model capable of describing strain relaxation under constant stress conditions (or zero stress) and stress relaxation under constant strain conditions. Besides volumetric strain relaxation during the fusion of two healthy cell aggregates, this model has been suitable for characterizing the long-time cell rearrangement of various multicellular systems such as: (1) cell aggregate rounding after uniaxial compression (Pajic-Lijakovic and Milivojevic, 2019a) formulated based on experimental data by Marmottant et al. (2009) and Mombash et al. (2005), (2) free expansion of cell monolayers (Pajic-Lijakovic and Milivojevic, 2020c) formulated based on the experimental data by Serra-Picamal (2012), and (3) the CCM of confluent monolayers (Pajic-Lijakovic and Milivojevic, 2020c) formulated based on the experimental data by Notbohm et al. (2016).

Otherwise, the aggregate system should be treated as a viscoelastic liquid. The simplest linear model for viscoelastic liquid is the Maxwell model (Pajic-Lijakovic, 2021). This model has been proposed for a long-time cell rearrangement in the form of weakly connected cell streams (Pajic-Lijakovic and Milivojevic, 2021a). The Maxwell model describes stress relaxation under constant strain rate, while the strain cannot relax (Pajic-Lijakovic, 2021). Guevorkian et al. (2011) confirmed the Maxwell model as suitable for describing the cell long-time rearrangement during the aggregate micropipette aspiration. The volumetric part of strain energy density is equal to: (1) $W_{VZ}(t) = \frac{1}{\Delta V}\int_{\Delta V} \frac{1}{2}\tilde{\sigma}_{RV}:\tilde{\varepsilon}_V \, d^3r$ for the Zener model and (2) $W_{VM}(t) = \frac{\Delta t}{\Delta V}\int_{\Delta V}\frac{1}{2}\tilde{\sigma}_{RV}:\dot{\tilde{\varepsilon}}_{cV} \, d^3r$ for the Maxwell model (where. $\tilde{\sigma}_{RV}$ is the local cell normal residual stress accumulation, $\tilde{\varepsilon}_V$ is the local volumetric strain, and $\dot{\tilde{\varepsilon}}_{cV}$ is the volumetric strain rate) (Pajic-Lijakovic and Milivojevic, 2021a).

An increase in the normal residual stress in the core regions of cell aggregates causes an increase in the cell packing density. This increase is induced by the volumetric cell rearrangement occurred through CCM as was shown in Figure 3a,b.

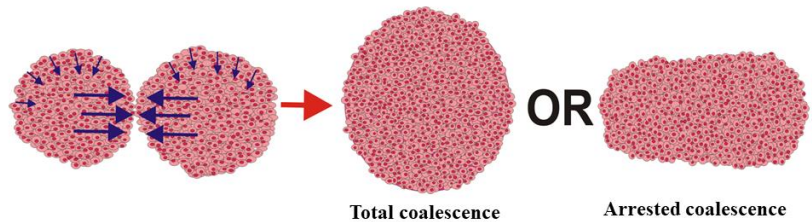

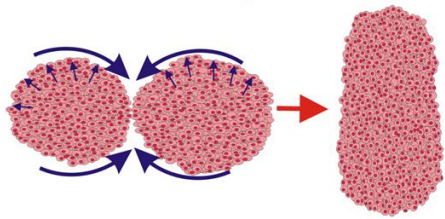

**Figure 3**. The fusion of two cell aggregates for (a) healthy epithelial cells and (b) cancer cells

CCM is intensive at the contact point between two cell aggregates which leads to an increase in the neck radius. In this case, generated two volumetric velocity fronts have an opposite directions (Figure 3a). Every cell velocity front is directed from the surface region of one aggregate to the core region of the other. Consequently, local collisions of velocity fronts induce an increase in cell packing density which can lead to the cell jamming state transition (Trepat et al., 2009; Nnetu et al., 2012; Pajic-Lijakovic and Milivojevic, 2019c;2021a). Local cell jamming state transitions near the contact point between two aggregates can induce damping effects of the aggregates fusion (Oriola et al., 2020). This phenomenon is known as the "arrested coalescence" (Oriola et al., 2020; Grosser et al., 2021). Consequently, the fusion of two healthy cell aggregates can follow two scenarios, i.e. the total coalescence or the arrested coalescence depending on (1) the fraction of cells in jamming state within the neck region and (2) the ability of cells to undergo unjamming transition (Figure 3a).

However, the fusion of two cancer cell aggregates sometimes follows different scenario, depending on the nature of cell-cell and cell-ECM contacts, as was shown in Figure 3b. The fusion also induces an increase in the neck radius while the surface and volume of two-aggregate system increase by forming the unregularly ellipsoidal shape. Dechristé et al. (2018) considered the fusion of two human carcinoma cell aggregates (HCT116 cell line) as a consequence of cell divisions within 70 h. The doubling time of HCT116 cells is 18 h (Gongora et al., 2008). Within this time period, cell divisions can be neglected, while the volume change occurs primarily via CCM. The results in the form of the volume ratio $\frac{V}{V_0}$ and surface ratio $\frac{A}{A_0}$ vs. time significantly depend on the initial aggregate size (Figure 4a,b). The volume ratio $\frac{V}{V_0}$ of larger aggregates (500 $\mu m$ diameter) increases 1.25 times, while the corresponding surface ratio $\frac{A}{A_0}$ increases 3.47 times within 18 h (where $V_0$ and $A_0$ are the initial volume and surface of the two-aggregate system, respectively and $V(t)$ and $A(t)$ are its current volume and surface) (Dechristé et al., 2018). The volume ratio $\frac{V}{V_0}$ of smaller aggregates (300 $\mu m$ diameter) increases 1.55 times, while the corresponding surface ratio $\frac{A}{A_0}$ increases 5.93 times within 18 h (Dechristé et al., 2018) (Figure 4a,b).

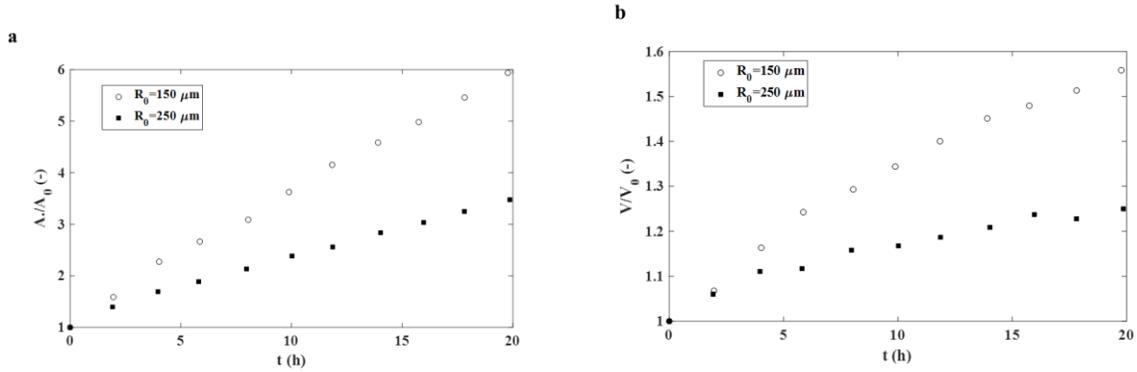

**Figure 4**. The fusion of HCT116 cell aggregates: (a) an increase in the surface of a two-aggregate system vs. time and (b) an increase in volume of a two-aggregate system vs. time.

The volume of larger two-aggregate system is capable of relaxing within 18 h, while the volume of smaller aggregate system cannot reach out the equilibrium state within this time period (Figure 4a). In both cases, the volume increase is limited. It is in accordance with the fact that this increase corresponds to a small decrease in the cell packing density in order to keep the system structural integrity. However, an increase in surface of two-aggregate system vs. time is more intensive and approximately linear (Figure 4b). The aspect ratio is $AR \approx 0.92$ for smaller aggregates after 18 h, which corresponds to a new ellipsoidal shape. The more intensive increase in the volume and surface of smaller aggregate system is caused by the larger surface-to-volume ratio. In contrast, the aspect ratio change is slower for larger aggregates and equal to $AR \approx 1.1$ after 18 h (Dechristé et al., 2018).

This result points to a surface rearrangement of cancer cells as a dominant scenario (Figure 3b). This surface rearrangement becomes more intensive with an increase in surface-to-volume ratio of two-aggregate system, i.e. for smaller cell aggregates. Grosser et al. (2021) considered and compared the fusion of two breast cell aggregates such as: (1) healthy epithelial MCF-10A cell lines and (2) cancerous mesenchymal MDA-MB-436 cells within 60 h. While MCF-10A cells undergo arrested coalescence and the average cell velocity drops to zero, cancer cells kept their velocity approximately constant within this time period and avoided the jamming state transition. This interesting result represents the confirmation of the surface activity of cancer cells. Instead of generating two opposite volumetric velocity fronts near the contact point between two aggregates, cancer cells rather undergo the surface cell rearrangement. The surface cell rearrangement occurs by generating the surface velocity fronts directed to the contact point between the aggregates (Figure 3b).

On that base, the rearrangement of cancer aggregates is accomplished without an increase in the cell packing density which could lead to the cell jamming state transition. This scenario of cell rearrangement during the aggregates fusion could be related to an invasiveness of cancer in the form of the surface velocity fronts.

Consequently, two driving forces guide the fusion of cell aggregates. One of them is the surface tension which acts to reduce the surface by inducing the volumetric cell rearrangement. The other driving force, related to activity of cancer cells, has an opposite rule. This driving force induces a cell surface rearrangement and leads to an increase in the surface of two-aggregate system. In order to deeply understand the nature of the long-time rearrangement of cancer cells, it is necessary to consider the surface tension as a product of an inter-relation between the cell contractility in one hand and the state of cell-cell and cell-ECM adhesion contacts in the other.

**Surface activity of cancerous cells: the state of cell-cell adhesion contacts**

The tissue surface tension is a product of cumulative effects of cell-cell interactions during CCM within the surface region of cell aggregates. Migrated cells use adherens junctions (AJs) to mechanically couple between each other and as an important source of signaling that coordinates collective behavior (Barriga and Mayor, 2019). One of the main components of AJs are proteins from the cadherin family. While epithelial cells form stronger E-cadherin based AJs, mesenchymal cells frequently form weak N-cadherin mediated AJs(Barriga and Mayor, 2019). The state of cell-cell adhesion contacts is influenced by cell contractility (Malinova and Huvanees, 2018) and the state of cell-ECM adhesion contacts (FAs) (Zuidema et al., 2020; Tian et al., 2021). Cadherin mediated AJs and integrin mediated FAs can have either an antagonistic or cooperative relation, which is mediated by their connection to the actin cytoskeleton and results in a balanced force distribution between sites of cell–cell and cell–ECM attachment (Zuidema et al., 2020). The establishing of integrin-mediated FAs stimulates cadherin accumulation which leads to an increase in the cell packing density (Lin et al., 2006). Tian et al. (2021) pointed out that the tension in FAs of the mesenchymal MDA-MB-231 cells increased on stiffer substrates while the tension in epithelial MCF-10A cells exhibited no apparent change against the substrate stiffness. This result represents a confirmation of cancer cell adaptability on the microenvironmental conditions.

Devanny et al. (2021) considered the long-time rearrangement and compaction of various breast cell aggregates such as: healthy epithelial MCF-10A cells and various mesenchymal cell lines, i.e. MDA-MB-468, along with MDA-MB-231 and MDA-MB-436 cells. They pointed out that the compaction of cancer cell aggregates occurs primarily via $\beta 1$ integrin-mediated adhesion complexes, while the compaction of MCF-10A cells occurs primarily via cadherin-mediated adhesion complexes.

Cell contractility influences the state of AJs and FAs as well as their crosstalk. Devanny et al. (2021) revealed that contractility plays a fundamentally different role in the cell lines in which compaction is driven primarily by integrins (MDA-MB-468, along with MDA-MB-231 and MDA-MB-436 cells) vs. by cadherins (MCF-10A). They considered the long-time rearrangement of healthy epithelial MCF-10A cell aggregates and emphasized that the cell contractility suppression reduces the tissue surface tension. Upon treatment with 20 μM blebbistatin, capable of suppressing cellular contractility, the aggregates lost the smooth surface and became more irregular in shape. In contrast, untreated MCF-10A cell aggregates display a smooth surface. Intensive contractility of surface cells enhances the strength of E-cadherin mediated adhesion contacts and on that base increases the tissue surface tension. Such increase leads to a cell movement from the aggregate surface region to its core region (Figure 3a). This movement results in a decrease in the aggregate surface and an increase in the normal residual stress accumulation within the aggregate core region as well as the cell packing density. An increase in cell packing density can induce the cell jamming state transition (Nnetu et al., 2012; Pajic-Lijakovic and Milivojevic, 2021a).

Enhanced contractility of cancer cells within the aggregate surface region induces an increase in cell-cell repulsions which reduce the tissue surface tension (Devanny et al., 2021). This repulsion has a feedback impact on the single-cell contractility itself. Warmt et al. (2021) reported that the contractility of breast mesenchymal cells is more intensive than the contractility of healthy breast epithelial MCF-10A cells. They revealed that intensive contractility of mesenchymal cells comes as a product of the stress fibers action. Otherwise, the contractility of MCF-10A cells represents the consequence of the contractility of cortical structure.

The reduction of the surface tension leads to a mesenchymal cell movement from the aggregate core region to its surface (Figure 3b) which results in an increase in the surface of two-aggregate system. This cell movement is capable of reducing undesirable accumulation of internal cell residual stress in the aggregate core region by decreasing the cell packing density (Trepat et al., 2009). This stress corresponds to a few kPa under physiological conditions and can reduce cell proliferation and movement (Kalli and Stylianopoulos, 2018). Dolega et al. (2017) reported that the cell normal residual stress accumulated in the core region of CT26 cancer cell aggregate (diameter of 240 $\mu m$) is about 8 times higher than stress at the aggregate surface under externally applied osmotic stress of 5 kPa. This osmotic stress corresponds to a physiological compressive stress occurs under in vivo conditions.

The surface rearrangement of cancer cells during the aggregate fusion is discussed based on the proposed mathematical model.

**Long-time cell rearrangement during the aggregate fusion: a modeling consideration**

The volume of two-aggregate system $V(t)$ can (1) increase for $\frac{V(t)}{V_0} > 1$ (for cancer cells), (2) stay constant for $\frac{V(t)}{V_0} = 1$ (for healthy epithelial cells), and (3) decrease for $\frac{V(t)}{V_0} < 1$ (for healthy epithelial cells) (Kosztin et al., 2012). The volume decrease is resulted by increasing in the cell packing density $\langle n(t) \rangle$, while the volume increase is connected with a decrease in the cell packing density, i.e. $\frac{d\langle n(t) \rangle}{dt} \sim \left(\frac{dV(t)}{dt}\right)^{-1}$. The volume of two-aggregate system can be expressed as:

$$V(t) = N_T \langle n(t) \rangle^{-1} \tag{3}$$

where $t$ is the time scale of hours, $N_T$ is the total number of cells such that $N_T \approx const$ during the aggregate fusion (the cell divisions are neglected), and $\langle n(t) \rangle$ is the average cell packing density equal to $\langle n(t) \rangle = \frac{1}{V(t)} \int n(r,t) d^3r$, $n(r,t)$ is the local cell packing density expressed as $n(r,t) = \sum_{i=1}^{N_T} \delta(r - r_i)$, and $\langle n(t) \rangle^{-1} = \langle V_c \rangle$ is the average volume per single cells equal to $\langle V_c \rangle = v_c + v_F$, $v_c$ is the single-cell volume, and $v_F$ is the free volume per single cells. The surface of two aggregate system $A(t)$ is equal to:

$$A(t) = N_s(t) \langle n_s(t) \rangle^{-1} \qquad (4)$$

where $N_s(t)$ is the number of cells located at the surface of aggregate system equal to $N_s(t) = \int n_s(\Re, t) d^2\Re$, $n_s(\Re, t)$ is the local cell surface packing density, and $\Re = \Re(x, y, z)$ is the coordinate of the surface. The $n_s(\Re, t)$ changes during the fusion process as a product of two opposite tendencies. The tissue surface tension acts to reduce the surface. However, the cell contractility for the integrin mediated cell-cell and cell-ECM adhesion contacts, characteristic for some types of cancer cells, induces repulsion between cells which cannot be compensated by weak cell-cell adhesion contacts (Devanny et al., 2021). Consequently, this repulsion among cells causes a decrease in the tissue surface tension which results in an increase in the surface of two-aggregate system. This repulsion leads to the enhanced effective surface activity of cancer cells. Accordingly, the change of the cell surface packing density $n_s(\Re, t)$ can be expressed based on particularly formulated phase model:

$$\frac{dn_s(\Re, t)}{dt} = \Gamma \left[ \frac{\delta F_s(n_s)}{\delta n_s} - k_B T_{eff} \ln(X_a(\Re, t)) \right] \qquad (5)$$

where $\Gamma$ is the kinetic constant and $F_s$ is the Landau-Ginzburg surface free energy (Cohen and Murray, 1981; Pajic-Lijakovic and Milivojevic, 2020a) equal to $F_s = \int \left[ f(n_s) + \frac{1}{2}\gamma(\nabla n_s)^2 + \cdots \right] d^2\Re$, $f(n_s)$ represents the energy density which this volume would have in homogeneous composition and the other terms represent the gradient energy contributions significant in non-homogeneous states while $\gamma$ is the surface tension which accounts for cumulative effects of change the state of adhesion contacts, $k_B$ is the Boltzmann constant, $T_{eff}$ is the effective temperature. The concept of effective temperature has been applied for considering a rearrangement of various thermodynamic systems from glasses and sheared fluids to granular systems (Casas-Vazquez and Jou, 2003). Pajic-Lijakovic and Milivojevic (2019c;2021a) applied this concept to a long-time rearrangement of dense cellular systems. The effective temperature, in this case, represents a product of cell mobility and has been expressed as $(k_B T_{eff})^{1/2} \sim \langle |\vec{v}_c| \rangle$ (where $\langle |\vec{v}_c| \rangle$ is the cell average speed). The cell activity ratio is expressed as $X_a(\Re, t) = \frac{a}{a^\theta}$ (where $a$ is the cell activity which satisfies the condition that $a \geq a^\theta$, while $a^\theta$ is the reference activity of health cells under physiological conditions). The first term of eq. 5 accounts for the surface tension action to increase the cell surface packing density and decrease the surface (eq. 4), while the second term represents the result of repulsion among cancer cells caused by their contractility which leads to a decrease in the cell surface packing density and consequently, an increase in the surface.

The activity of cancer cells as well as their contractility depends on: (1) cell mobility expressed by the effective temperature, (2) cell surface packing density, (3) the cumulative energy of AJs and FAs equal to $F_A = \sum_j \mu_j \delta(\Re - \Re_j)$ (where $\mu_j$ is the chemical potential of the j-th adhesion contact). The change of $X_a$ can be expressed as:

$$\frac{d X_a(\Re, t)}{dt} = \left(\frac{\partial X_a}{\partial T_{eff}}\right)_{n_s, F_A} \frac{\partial T_{eff}}{\partial t} - \left(\frac{\partial X_a}{\partial n_s}\right)_{T_{eff}, F_A} \frac{\partial n_s}{\partial t} - \left(\frac{\partial X_a}{\partial T_{eff}}\right)_{n_s, T_{eff}} \frac{\partial F_A}{\partial t} \qquad (6)$$

The cell activity increases with an effective temperature, while an increase in the cell surface packing density and the strength of cell-cell adhesion contacts reduces the cell activity as well as the cell invasiveness. An increase in $n_s$ under constant number of cells within the surface $N_s$ leads to a decrease in the surface of two-aggregate system $A(t)$ during the fusion (eq. 4).

**Conclusion**

The surface activity of mesenchymal cells is estimated by considering the simple model system such as the fusion of two cell aggregates. While the fusion of epithelial cell aggregate, driven by the tissue surface tension, leads to a decrease in the surface and volume of two-aggregate systems, the fusion of mesenchymal cell aggregates follows quite different scenario. In this case, an increase in the surface of two-aggregate system is intensive, while an increase in its volume is limited. While the epithelial cells perform volumetric cell rearrangement induced by decreasing the surface, cancer cells undergo surface rearrangement and on that base avoid the cell jamming state transition. The origin of this interesting phenomenon lays in ability of cancer cells to reduce the tissue surface tension and behave as surface active constituents.

The surface activity of mesenchymal cells is closely connected with the crosstalk between cell-cell and cell-ECM adhesion complexes obtained for contractile cells. While epithelial cells establish stronger E-cadherin mediated cell-cell adhesion contacts, mesenchymal cells form weak cell-cell adhesion contacts and $\beta 1$ integrin-mediated cell-ECM adhesion contacts. Cellular contractions, significant in the aggregate surface region, generate repulsion among cancer cells which results in a decrease in the tissue surface tension. In contrast, contractility of epithelial cells leads to establishing stronger cell-cell adhesion contacts which leads to an increase in the tissue surface tension. Change the state of cell-cell adhesion contacts influenced by the cellular contractility has a feedback impact on the contractility itself.

The surface activity of cancer cells depends on: (1) the mobility of cancer cells – modeled by the effective temperature, (2) strength of cell-cell adhesion contacts –modeled by proper free energy functional, and (3) the cell surface packing density. These factors obtained at supracellular level influence the single-cell contractility and on that base the surface activity of single cells.

Additional experiments are necessary in order to: (1) estimate the molecular mechanisms of change in the state of cell-cell adhesion contacts influenced by cellular contractions for cancer mesenchymal cells and healthy epithelial cells, (2) consider the feedback impact of these changes on cellular contractility itself, and (2) correlate the collective effects of these changes with the tissue surface tension and invasiveness of cancer cells.

**Acknowledgement**. This work was supported by the Ministry of Education, Science and Technological Development of the Republic of Serbia (Contract No. 451-03-9/2021-14/200135).

**Conflict of interest:** We have no conflict of interest.

# References

1. Barriga EH, Mayor R. 2019. Adjustable viscoelasticity allows for efficient collective cell migration. Sem Cell Dev Biol 93:55-68.

2. Beunk L, Brown K, Nagtegaal I, Friedl P, Wolf K. 2019. Cancer invasion into musculature: Mechanics, molecules and implications. Sem Cell Dev Biol. 93, 36-45.

3. Blanchard GB, Fletcher AG, Schumacher LJ. 2019. The devil is in the mesoscale: mechanical and behavioural heterogeneity in collective cell movement. Sem Cell Dev Biol 93:46-54.

4. Casas-Vazquez J, Jou D. 2003. Temperature in non-equilibrium states: a review of open problems and current proposals. Rep Prog Phys 66:1937-2023.

5. Clark AG and Vignjevic DM. 2015. Models of cancer cell invasion and the rule of microenvironment, Cur. Op. Cell Biol. 36:13-22.

6. Cohen DS and Murray JD. 1981. A Generalized Diffusion Model for Growth and Dispersal in a Population. J. Mat. Biol. 12: 237-249.

7. Dechristé G, Fehrenbach J, Griseti E, Lobjois V, Clair Poignar C. 2018. Viscoelastic modeling of the fusion of multicellular tumor spheroids in growth phase. J Theor Biol 454:102–109.

8. Devanny AJ, Vancura MB, Kaufman LJ. 2021. Exploiting Differential Effects of Actomyosin Contractility to Control Cell Sorting Among Breast Cancer Cells. Mol Biol Cell doi.org/10.1091/mbc.E21-07-0357.

9. Dolega, M.E., Delarue, M., Ingremeau, F., Prost, J., Delon, A., Cappello, G. 2017. Cell-like pressure sensors reveal increase of mechanical stress towards the core of multicellular spheroids under compression. Nature Comm. 8, 14056 1-9.

10. Gandalovičová A, Vomastek T, Rosel D, Brábek J. 2016. Cell polarity signaling in the plasticity of cancer cell invasiveness. Oncotarget 7(18):25022-25049.

11. Gongora C, Candeil L, Vezzio N, Copois V, Denis V, Breil C, Molina F, Fraslon C, Conseiller E, et al. 2008. Altered expression of cell proliferation-related and interferon-stimulated genes in colon cancer cells resistant to SN38. Cancer Biol & Therapy 7(6):822-32.

12. Grosser S, Lippoldt J, Oswald L, Merkel M, Sussman DM, Renner F, Gottheil P, Morawetz EW, Fuhs T, Xie X, et al. 2021. Cell and Nucleus Shape as an Indicator of Tissue Fluidity in Carcinoma. Phys Rev X 11, 011033.

13. Guevorkian K, Gonzalez-Rodriguez D, Carlier C, Dufour S, Brochard-Wyart F. 2011. Mechanosensitive shivering of model tissues under controlled aspiration. PNAS 108(33):13387-13392.

14. Kalli, M., Stylianopoulos, T. 2018. Defining the Role of Solid Stress and Matrix Stiffness in Cancer Cell Proliferation and Metastasis. Frontiers in Oncology doi: 10.3389/fonc.2018.00055.

15. Kosztin I, Vunjak-Novakovic G, Forgacs G. 2012. Colloquium: Modeling the dynamics of multicellular systems: Application to tissue engineering. Rev Mod Phys 84(4):1791-1805.

16. Kubitschke H, Blauth E, Gottheil P, Grosser S, Kaes J. 2021. Jamming in Embryogenesis and Cancer Progression. Front Phys doi: 10.3389/fphy.2021.666709.

17. Lin, RZ, Chou, LF, Chien, CCM, and Chang, HY. 2006. Dynamic analysis of hepatoma spheroid formation: Roles of E-cadherin and β1-integrin. Cell Tissue Res 324, 411–422.

18. Malinova TS, Huveneers S. 2018. Sensing of Cytoskeletal Forces by Asymmetric Adherens Junctions. Trends Cell Biol 28(4):328-341.

19. Marmottant P, Mgharbel A, Kafer J, Audren B, Rieu JP, Vial JC, van der Sanden B, Maree AFM, Graner F, Delanoe-Ayari H. 2009. The role of fluctuations and stress on the effective viscosity of cell aggregates. PNAS 106(41):17271-17275.

20. Mombach JCM, Robert D, Graner F, Gillet G, Thomas GL, Idiart M, Rieu JP. 2005. Rounding of aggregates of biological cells: Experiments and simulations. Phys A 352:525-534.

21. Murray JD, Maini PK, Tranquillo RT. 1988. Mechanochemical models for generating biological pattern and form in development. Physics Report 171(2):59-84.

22. Nnetu KD, Knorr M, Kaes J, Zink M. 2012. The impact of jamming on boundaries of collectively moving weak-interacting cells. New J Phys 14:115012.

23. Notbohm J, Banerjee S, Utuje KJC, Gweon B, Jang H, Park Y, Shin J, Butler JP, Fredberg JJ, Marchetti MC. 2016. Cellular Contraction and Polarization Drive Collective Cellular Motion, Biophys J 110 (12): 2729-2738.

24. Oriola D, Marin-Riera M, Anlas K, Gritti N, Matsumiya M, Aalderink G, Ebisuya M, Sharpe J, Trivedi V. 2020. Arrested coalescence of multicellular aggregates. doi:https://arxiv.org/abs/2012.01455.

25. Pajic-Lijakovic I. and Milivojevic M. 2019a. Long-time viscoelasticity of multicellular surfaces caused by collective cell migration – multi-scale modeling considerations. Sem Cell Dev Biol 93:87-96.

26. Pajic-Lijakovic I. and Milivojevic M. 2019b. Functional epithelium remodeling in response to applied stress under in vitro conditions. Appl Bionics Biomech. doi.org/10.1155/2019/4892709.

27. Pajic-Lijakovic I. and Milivojevic M. 2019c. Jamming state transition and collective cell migration. J Biol Eng 13: 73 doi.org/10.1186/s13036-019-0201-4.

28. Pajic-Lijakovic, I., Milivojevic, M., 2020a. Viscoelasticity of multicellular systems caused by collective cell migration: dynamics at the biointerface. Europ Biophys J 49:253-265.

29. Pajic-Lijakovic I. and Milivojevic M. 2020c. Mechanical oscillations in 2D collective cell migration: the elastic turbulence. Front Phys doi: 10.3389/fphy.2020.585681.

30. Pajic-Lijakovic I. and Milivojevic M. 2021a. Multiscale nature of cell rearrangement caused by collective cell migration. Europ Biophys J 50:1-14.

31. Pajic-Lijakovic I. and Milivojevic M. 2021b. Viscoelasticity and cell jamming state transition. Europ Phys J Plus 136:750, doi.org/10.1140/epjp/s13360-021-01730-3

32. Pajic-Lijakovic I. 2021. Basic concept of viscoelasticity, in Viscoelasticity and collective cell migration, eds. I. Pajic-Lijakovic and E. Barriga, Chapter 2, (Academic Press, United States, 2021), p. 21.

33. Serra-Picamal X, Conte V, Vincent R, Anon E, Tambe DT, Bazellieres E, Butler JP, Fredberg JJ, Trepat X. 2012. Mechanical waves during tissue expansion, Nature Phys 8 (8):628-634.

34. Petrungaro, G., Morelli, L., Uriu, K., 2019. Information flow in the presence of cell mixing and signaling delays during embryonic development. Sem Cell Dev Biol 93:26-35.

35. Shafiee A, McCune M, Forgacs G, Kosztin I. 2015. Post-deposition bioink self-assembly: a quantitative study. Biofabric 7:045005, doi:10.1088/1758-5090/7/4/045005.

36. Tian F, Lin TC, Wang L, Chen S, Chen X, Yiu PM, Tsui OKC, Chu J, Kiang CH, Park H. 2021. Mechanical Responses of Breast Cancer Cells to Substrates of Varying Stiffness Revealed by Single-Cell Measurements. J. Phys. Chem. Lett. 11:7643−7649.

37. Trepat X, Wasserman MR, Angelini TE, Millet E, Weitz DA, Butler JP, Fredberg JJ. 2009. Physical forces during collective cell migration. Nature Phys 5:426-430.

38. Warmt E, Grosser S, Blauth E, Xie X, Kubitschke H, Stange R, Sauer F, Schnaub J, Tomm JM, von Bergen M, Kaes J. 2021. Differences in cortical contractile properties between healthy epithelial and cancerous mesenchymal cells. New J Phys doi.org/10.1088/1367-2630/ac254e.

39. Zuidema A, Wang W, Sonnenberg A. 2020. Crosstalk between Cell Adhesion Complexes in Regulation of Mechanotransduction. BioAssey 42:2000119, DOI: 10.1002/bies.202000119.
18